\documentclass[reprint,amsmath,amssymb,aps,prl,showpacs]{revtex4-1}

\usepackage{graphicx}
\usepackage{dcolumn}
\usepackage{bm}

\begin{document}

\title{Vacuum state truncation via the quantum Zeno effect}
\author{Tae-Gon Noh}
\affiliation{Department of Physics, Kyungsung University, Busan 608-736, Korea}

\date{August 20, 2012}

\begin{abstract}
In the context of quantum state engineering we analyze the effect of observation on nonlinear optical
$n$-photon Fock state generation. We show that it is possible to truncate the vacuum component from an
arbitrary photon number superposition without modifying its remaining parts.  In the course of the full
dynamical analysis of the effect of observation, it is also found that the Zeno and the anti-Zeno effects
repeat periodically. We discuss the close relationship between vacuum state truncation and so-called
``interaction-free'' measurement.
\end{abstract}


\maketitle

The engineering of quantum states of an electromagnetic field is essential to fundamental studies of quantum
optics \cite{Gerry05, Dell'Anno06}, and is also important for the development of optical quantum information
technologies \cite{Nielsen00, Gisin02, Kok07}. The quantum-optical state engineering method we are interested
in here is optical state truncation. Pegg, Phillips, and Barnett presented a scheme to prepare any required
superposition of the vacuum and one-photon states by the choice of a proper measurement after a beam
splitter. Because their scheme is effectively the same as a physical truncation of the photon number
superposition making up the coherent state involved during the measurement, they called the device ``quantum
scissors'' \cite{Pegg98, Leonski11}.

In this Letter, we present a novel scheme of quantum scissors in which only the vacuum component is truncated
from any given photon number superposition state without corrupting the remaining parts. The vacuum state has
a unique property different from other photon number states, in that it represents the ``absence'' of a
photon while the other states (or their superpositions) represent the ``presence'' of photons. Considering a
physical measurement using the electromagnetic field as a probe of the measurement, the absence of a photon
may in a sense correspond to the absence of measurement. Therefore, one can expect that the presence or
absence of photons may lead to totally different physical effects: ``measurement'' or ``no measurement''. As
a specific example, we will consider the quantum Zeno effect, which refers to the suppression of the free
dynamical evolution of a system when the system is subject to repeated measurements \cite{Misra77, Peres80,
Luis96, Facchi08}. In the limit of infinitely frequent or continuous measurements, the observed system would
be locked to its initial state. If the initial probe state is a superposition of the presence and absence of
photons, then we may expect the total output state of the system and probe to be a superposition of two
different physical effects: the occurrence and nonoccurrence of the Zeno effect. Thus, we show that it is
possible to effectively truncate the vacuum component of the probe state by projecting the output state onto
the state of occurrence of the Zeno effect.

The present scheme is also closely related to the so-called ``interaction-free'' measurement (IFM) (also
known as ``quantum interrogation''), which refers to the determination of the presence of an object without
any photons being scattered by the object, thereby excluding any disturbance of it \cite{Renninger60,
Dicke81,  EV93, Kwiat95, Jozsa99, Hosten06, Vaidman07, Noh09}. If we consider a coherent superposition of the
presence and absence (the vacuum) of a quantum object, the IFM of the object becomes equivalent to vacuum
state truncation. Thus, our scheme may be considered a new scheme of IFM in which the inferred object is an
electromagnetic field with an IFM efficiency approaching 100\%.

We give a detailed description of the vacuum state truncation in the following. We first consider Kilin and
Horoshko (KH)'s scheme for the generation of a pure Fock state of a single-mode field via a single-pass
parametric interaction in a nonlinear medium \cite{Kilin95}. In the KH scheme, the nonlinear medium is pumped
by a classical coherent field to amplify the signal mode from the vacuum state into the $n$-photon Fock state
by the $(n+2)$th order nonlinear interaction. The interaction Hamiltonian describing this parametric process
is
\begin{eqnarray}
H_n = \frac{\hbar g}{\sqrt{n!}}
       \Big[ \big( a^n +a^{\dagger n} \big)
       - \frac{1}{n} \big( a^\dagger a^{n+1} + a^{\dagger n+1} a \big) \Big],
\label{eq:KH}
\end{eqnarray}
where $a^\dagger$ ($a$) is the photon creation (annihilation) operator of the signal mode, and $g$ is a real
coupling parameter that is related to the nonlinear susceptibilities satisfying the condition: $\chi^{(n)} =
- n \chi^{(n+2)}$. By means of the interaction Hamiltonian Eq.~(\ref{eq:KH}), the initial vacuum state
evolves as
\begin{eqnarray}
U_{KH}(\tau) |0\rangle_a &=&  \exp ( i H_n \tau/\hbar ) |0\rangle_a  \nonumber \\
                       &=& \cos (g \tau) |0\rangle_a + i \sin (g \tau) |n\rangle_a.
\end{eqnarray}
If the interaction time is chosen as $\tau = \pi/2g$, the output state becomes a pure $n$-photon Fock state.

We now discuss the manifestation of the Zeno effect in the KH scheme. Suppose the system is subject to
repeated measurements at each very short time interval $\Delta \tau = \tau /N$ with large $N$. After the
first measurement, the probability of finding the signal mode in the $n$-photon state is $\sin^2 (g
\Delta\tau) \approx (g \Delta\tau)^2$, which is very small for large $N$. The probability of remaining in the
initial vacuum state is $\cos^2 (g \Delta\tau) \approx 1 - (g \Delta\tau)^2$. The signal field is found in
the vacuum state with relatively large probability at each measurement stage. Thus, the probability that the
signal mode is found in the initial state after all $N$ repeated measurements is
\begin{eqnarray}
[ \cos^2 (g \Delta\tau) ]^N \approx 1 - (g \tau)^2 /N.
\end{eqnarray}
The signal mode would be locked to the initial vacuum state with certainty in the limit $N \rightarrow
\infty$, i.e., each measurement reduces the signal field into $|0\rangle_a$ and the Zeno effect occurs.

\begin{figure}
\includegraphics[width=7.0cm]{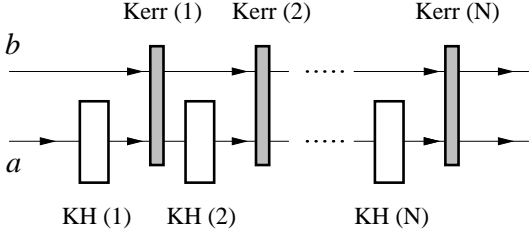}
\caption{\label{fig:Fig1} Proposed scheme of vacuum state truncation.}
\end{figure}

For a full dynamical analysis of the Zeno effect, we consider a modification of the KH scheme as depicted in
Fig.~\ref{fig:Fig1}. The KH's nonlinear medium is now divided into $N$ equal pieces with the interaction time
$\Delta \tau = \tau /N = \pi/2gN$ within each part. Assume that if the pieces are perfectly aligned, and if
the field evolution is not disturbed after each piece, then the input vacuum state $|0\rangle_a$ in mode $a$
would coherently evolve into the $n$-photon Fock state $|n\rangle_a$ as in the original KH scheme. However,
in Fig.~\ref{fig:Fig1}, we employ the optical Kerr effect to implement quantum nondemolition (QND)
measurement of the photon number in the mode $a$ \cite{Braginsky80, Imoto85, Gerry08}. The  modes $a$ and $b$
are coupled with each other by a series of $N$ nonlinear Kerr mediums. As is well known in the QND scheme
using the optical Kerr effect \cite{Imoto85}, the field of two coupled modes experiences a phase shift
proportional to the photon number of another field. One can determine the photon number without destroying
photons by the measurement of a phase shift imposed on another field. Thus, the field evolution of KH type in
mode $a$ would be modified by the dispersive coupling with mode $b$.

The unitary operator describing the state evolution within each of the Kerr mediums is given by
\begin{eqnarray}
U_{K} = \exp (i\kappa a^\dagger a b^\dagger b),
\end{eqnarray}
where $a^\dagger$ and $b^\dagger$ ($a$ and $b$) are the creation (annihilation) operators of the
corresponding modes, and $\kappa$ is a coupling parameter characterizing the Kerr interaction strength. For
convenience, we assume the same interaction strength for each Kerr medium, and the self-phase modulation
effect is neglected. We suppose the input field state in mode $b$ is an arbitrary state $|\Phi\rangle_b$,
which can be expanded in terms of the number states as
\begin{eqnarray}
|\Phi\rangle_b = \sum_{m=0}^{\infty} \alpha_m |m\rangle_b.
\label{eq:5}
\end{eqnarray}
Then, the total initial state of modes $a$ and $b$ is
\begin{eqnarray}
|\Psi^{(0)}\rangle = |0\rangle_a |\Phi\rangle_b.
\end{eqnarray}
After the first KH and Kerr medium, the state evolves as
\begin{eqnarray}
|\Psi^{(1)}\rangle
     &=& U_{K}U_{KH}(\Delta \tau) |\Psi^{(0)}\rangle  \nonumber \\
     &=&  \big[ \cos (g \Delta \tau) |0\rangle_a
           + i e^{i\kappa n b^\dagger b} \sin (g \Delta \tau) |n\rangle_a \big] |\Phi\rangle_b. \nonumber \\
\label{eq:evolve1}
\end{eqnarray}
We note that when the input state in mode $b$ is a number state $|\Phi\rangle_b = |m\rangle_b$, the phase
factor becomes $e^{i\kappa nm}$. This phase factor will vanish for $\kappa = 2\pi l/nm$ ($l= 0, 1, 2,
\cdots$), which means that the optical Kerr effect is effectively removed in this case and the original KH
process is recovered. Because our aim is to find the effect of any possible disturbance imposed by the QND
measurement, we discard this case. That is, we suppose the coupling parameter is chosen appropriately so that
$\kappa \neq 2\pi l/nm$.

Thus, this process is repeated $N-1$ more times as shown in Fig.~\ref{fig:Fig1}. After some calculation, we
obtain the final output state
\begin{eqnarray}
|\Psi^{(N)} \rangle
       &=& U_{K} U_{KH}(\Delta\tau) |\Psi^{(N-1)} \rangle  \nonumber \\
       &=& \big[ v(b^\dagger b) |0\rangle_a + w(b^\dagger b) |n\rangle_a \big] |\Phi\rangle_b.
\end{eqnarray}
Here, $v(b^\dagger b)$ and $w(b^\dagger b)$ are given by
\begin{eqnarray}
v(b^\dagger b) &=& e^{iN\delta/2}\bigg[ -i \bigg( \frac{\cos(g \Delta \tau)
                                \sin(\delta/2)}{\sin(\eta)} \bigg) \sin(N\eta)    \nonumber \\
               &&  + \cos(N\eta) \bigg],     \\
w(b^\dagger b) &=& i e^{i(N+1)\delta/2}
                    \bigg( \frac{\sin(g \Delta \tau)}{\sin(\eta)} \bigg) \sin(N\eta),
\end{eqnarray}
where $\delta = \kappa n b^\dagger b$, and $\eta = \cos^{-1} [ \cos(g \Delta \tau) \cos (\delta/2) ]$.
Using Eq.~(\ref{eq:5}), the output state is written as
\begin{eqnarray}
|\Psi^{(N)} \rangle
       &=& \alpha_0 \big[ \cos (N g \Delta \tau) |0\rangle_a
            + i \sin (N g \Delta \tau) |n\rangle_a \big] |0\rangle_b  \nonumber \\
       & & + \big[ v(b^\dagger b) |0\rangle_a
                      + w(b^\dagger b) |n\rangle_a \big] |\Phi^\prime \rangle_b,
\label{eq:evolve2}
\end{eqnarray}
where the state $|\Phi^\prime \rangle_b$ does not include the vacuum component, i.e.,
\begin{eqnarray}
|\Phi^\prime \rangle_b = \sum_{m=1}^{\infty} \alpha_m |m\rangle_b = |\Phi\rangle_b - \alpha_0 |0\rangle_b.
\end{eqnarray}
Since $g \Delta \tau= \pi/2N$, the first term of Eq.~(\ref{eq:evolve2}) becomes $i \alpha_0 |n\rangle_a
|0\rangle_b$, and it corresponds to the isolated dynamics of KH's procedure. The other term exhibits some
modifications of the KH process induced by the photons in mode $b$.

In the limit of large $N$, $v(b^\dagger b)$ and $w(b^\dagger b)$ become
\begin{eqnarray}
v(b^\dagger b) &=& 1 - \bigg( \frac{ i\pi^2 }{ 8 \tan(\delta/2) } \bigg)
                      \frac{1}{N} + O( N^{-2} ),           \\
w(b^\dagger b) &=& i e^{i(N+1)\delta/2}
                   \bigg( \frac{ \pi\sin(N\delta/2) }{ 2\sin(\delta/2) } \bigg)
                      \frac{1}{N} + O( N^{-2} ).           \nonumber \\
\label{eq:14}
\end{eqnarray}
Therefore, we obtain
\begin{eqnarray}
|\Psi^{(N)} \rangle \rightarrow
                     i \alpha_0 |n\rangle_a |0\rangle_b + |0\rangle_a |\Phi^\prime \rangle_b
             ~~~~\mbox{as}~    N \rightarrow \infty.
\label{eq:evolve3}
\end{eqnarray}
This entangled state represents the desired correlation between modes $a$ and $b$. For a finite number $N$,
the state $|\Phi^\prime \rangle_b$ is disturbed as a result of the measurements, although the photon number
is conserved in the QND measurements. However, Eq.~(\ref{eq:evolve3}) indicates that the disturbance imposed
on $|\Phi^\prime \rangle_b$ becomes arbitrarily small in the limit of very frequent measurements. That is,
the quantum state of the $|\Phi^\prime \rangle_b$ component in mode $b$ remains in its initial state while
the mode $b$ is correlated with mode $a$. In terms of quantum scissors, by projecting the output state
Eq.~(\ref{eq:evolve3}) onto the state $|0\rangle_a$, one can truncate only the vacuum component $|0\rangle_b$
in mode $b$ without corrupting the remaining parts $|\Phi^\prime \rangle_b$. Note also that if the mode $b$
is initially in the pure $|\Phi^\prime \rangle_b$ state (i.e., $\alpha_0 = 0$), then Fock state generation in
mode $a$ is always inhibited in the limit $N \rightarrow \infty$. Both the modes $a$ and $b$ are locked to
their initial states in this limit, and the Zeno effect occurs.

\begin{figure}
\includegraphics[width=8.5cm]{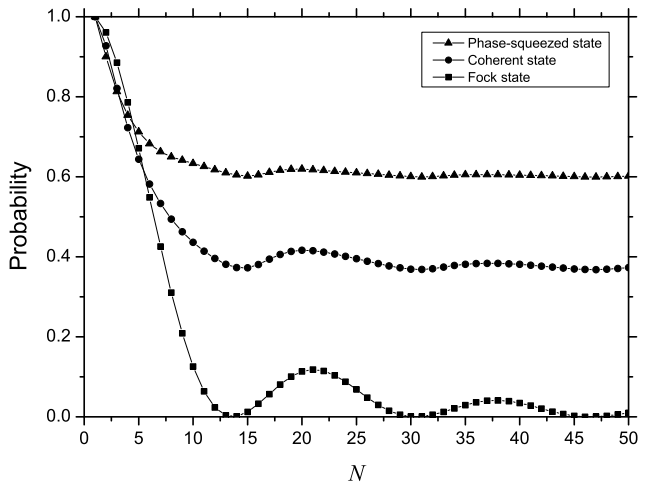}
\caption{\label{fig:Fig2} Photon emission probabilities ($P_{n=2}$) 
as a function of $N$. The input state of the mode $b$ is a 1-photon
Fock state (squares), a coherent state (circles), and a
phase-squeezed state (triangles). The probabilities approach the
value $|\alpha_0|^2$ as $N$ becomes large (Eq.~(\ref{eq:evolve3})).}
\end{figure}

In Fig.~\ref{fig:Fig2} we verify our results by representing the probability $P_n$ of $n$-photon Fock state
generation in mode $a$ as a function of $N$. For an arbitrary state $|\Phi\rangle_b$, the probability becomes
\begin{equation}
P_n = |\alpha_0 |^2 + |w(b^\dagger b) |\Phi^\prime \rangle_b |^2
    = |\alpha_0 |^2 + \sum_{m=1}^{\infty} |\alpha_m w(m)|^2,
\label{eq:prob}
\end{equation}
which is always less than or equal to 1. We plot the probability using a numerical calculation of
Eq.~(\ref{eq:prob}) for when the input state of the mode $b$ is a 1-photon Fock state, a coherent state, and
a phase-squeezed state, respectively. We suppose the states have the same mean photon number of $\langle m
\rangle_b = 1$. For the case of the phase-squeezed state
\begin{equation}
|\Phi\rangle_b = e^{ \alpha b^\dagger - \alpha^\ast b }
                 e^{ \frac{1}{2} \varepsilon^\ast b^2
                - \frac{1}{2} \varepsilon b^{\dagger 2} } |0\rangle_b,
\end{equation}
we choose $\varepsilon = -0.5$ and $\alpha = 0.853498$ to make $\langle m \rangle_b = 1$. We also suppose
$n=2$, i.e., 2-photon Fock state generation in mode $a$, and the Kerr interaction strength $\kappa = 0.2$.

As expected, Fig.~\ref{fig:Fig2} shows that $P_{n=2} \rightarrow |\alpha_0 |^2$ in each case as $N
\rightarrow \infty$. In addition, interestingly, it shows oscillation, i.e., the inhibition and enhancement
of photon emission in mode $a$ is repeated. The enhancement of photon emission by measurement is known as the
inverse or anti-Zeno effect \cite{Schieve89}, and has been discussed in parametric down-conversion
\cite{Luis98}. Using the arguments in \cite{Luis98}, the oscillatory behavior is attributed to the
interference of photon emission probability amplitudes originating from each part of the KH's nonlinear
medium. This interference becomes partially destructive (inhibition) or constructive (enhancement) according
to the phase mismatch caused by the Kerr interaction. We also note that the Mandel $Q$-parameter is $Q =
(\langle\Delta^2 m \rangle_b - \langle m \rangle_b ) / \langle m \rangle_b = -1$ for the 1-photon Fock state,
$Q = 0$ for the coherent state, and $Q = 1.67071$ for the super-Poissonian phase-squeezed state. Obviously,
the amplitude of the oscillation is reduced as the $Q$-parameter increases. This is because the photon number
fluctuation in mode $b$ results in the fluctuation of phase mismatch. Thus, in Fig.~\ref{fig:Fig2}, the
1-photon Fock state mode $b$ shows the greatest oscillation. From Eq.~(\ref{eq:14}), it can be seen that if
the input state in mode $b$ is a pure $m$-photon Fock state, the period of oscillation is $N \approx 2\pi /
\delta_0$, with $\delta_0$ defined by $\delta_0 =\kappa nm - 2\pi l$ ($0 \leq \delta_0 < 2\pi$). On the other
hand, the coherence between probability amplitudes decreases as $N$ increases, i.e., the photon emission
events in each part of the nonlinear medium become more distinguishable as the accuracy of the measurement
increases. Therefore, the overall behavior is a gradual suppression of photon emission.

In summary, we have presented a scheme of vacuum state truncation from an arbitrary photon number
superposition without modifying its remaining parts. The scheme is a direct consequence of the quantum
Zeno effect in nonlinear optical $n$-photon Fock state generation. During the full dynamical analysis of
the effect, we have also found the observation-induced oscillation in photon emission probabilities. The
proposed scheme can be considered a novel IFM in which the inferred object is a light field without the
vacuum component (i.e., $\sum_{m=1}^{\infty} \alpha_m |m\rangle_b$). It is interesting to note that one
cannot obtain any further information about the object details by vacuum state truncation (or IFM): all
one knows is that some state orthogonal to the vacuum is present.


\begin{thebibliography}{99}
\bibitem{Gerry05} C. C. Gerry and P. L. Knight, {\it Introductory Quantum Optics}
                   (Cambridge Univ. Press, 2005).
\bibitem{Dell'Anno06} F. Dell'Anno, S. De Siena, and F. Illuminati, Phys. Rep. {\bf 428}, 53 (2006).
\bibitem{Nielsen00} M. A. Nielsen and I. L. Chuang, {\it Quantum Computation and Quantum Information}
                    (Cambridge Univ. Press, 2000).
\bibitem{Gisin02} N. Gisin, G. Ribordy, W. Tittel, and H. Zbinden, Rev. Mod. Phys. {\bf 74}, 145 (2002).
\bibitem{Kok07} P. Kok, W. J. Munro, K. Nemoto, T. C. Ralph, J. P. Dowling, and G. J. Milburn,
                 Rev. Mod. Phys. {\bf 79}, 135 (2007).
\bibitem{Pegg98} D. T. Pegg, L. S. Phillips, and S. M. Barnett, Phys. Rev. Lett. {\bf 81}, 1604 (1998).
\bibitem{Leonski11} W. Leo\'{n}ski and A. Kowalewska-Kud{\l}aszyk, in {\it Progress in Optics},
edited by E. Wolf (Elsevier, 2011), Vol. 56, pp. 131-185.
\bibitem{Misra77} B. Misra and E. C. G. Sudarshan, J. Math. Phys. {\bf 18}, 756 (1977).
\bibitem{Peres80} A. Peres, Am. J. Phys. {\bf 48}, 931 (1980).
\bibitem{Luis96} A. Luis and J. Pe$\check{{\rm r}}$ina, Phys. Rev. Lett. {\bf 76}, 4340 (1996).
\bibitem{Facchi08} P. Facchi and S. Pascazio, J. Phys. A {\bf 41}, 493001 (2008).
\bibitem{Renninger60} M. Renninger, Z. Phys. {\bf 158}, 417 (1960).
\bibitem{Dicke81} R. H. Dicke, Am. J. Phys. {\bf 49}, 925 (1981).
\bibitem{EV93} A. C. Elitzur and L. Vaidman, Found. Phys. {\bf 23}, 987 (1993).
\bibitem{Kwiat95} P. Kwiat, H. Weinfurter, T. Herzog, A. Zeilinger, and M. A. Kasevich,
                  Phys. Rev. Lett. {\bf 74}, 4763 (1995).
\bibitem{Jozsa99} R. Jozsa, in {\it Lecture Notes in Computer Science},
                  edited by C. P. Williams (Springer-Verlag, Berlin, 1999), Vol. 1509, pp. 103-112.
\bibitem{Hosten06} O. Hosten, M. T. Rakher, J. T. Barreiro, N. A. Peters, and P. G. Kwiat,
                   Nature (London) 439, 949 (2006).
\bibitem{Vaidman07} L. Vaidman, Phys. Rev. Lett. {\bf 98}, 160403 (2007).
\bibitem{Noh09} T.-G. Noh, Phys. Rev. Lett. {\bf 103}, 230501 (2009).
\bibitem{Kilin95} S. Ya. Kilin and D. B. Horoshko, Phys. Rev. Lett. {\bf 74}, 5206 (1995).
\bibitem{Braginsky80} V. B. Braginsky, Y. I. Vorontsov, and K. S. Thorne, Science {\bf 209}, 547 (1980).
\bibitem{Imoto85} N. Imoto, H. A. Haus, and Y. Yamamoto, Phys. Rev. A {\bf 32}, 2287 (1985).
\bibitem{Gerry08} C. C. Gerry and T. Bui, Phys. Lett. A {\bf 372}, 7101 (2008).
\bibitem{Schieve89} W. C. Schieve, L. P. Horwitz, and J. Levitan, Phys. Lett. A {\bf 136}, 264 (1989).
\bibitem{Luis98} A. Luis and L. L. S$\acute{{\rm a}}$nchez-Soto, Phys. Rev. A {\bf 57}, 781 (1998).
\end{thebibliography}
\end{document}